\documentclass[12pt]{article}
\usepackage{epsfig}

\newcommand{\eq}{\begin{equation}}
\newcommand{\eqx}{\end{equation}}
\newcommand{\eqn}{\begin{eqnarray}}
\newcommand{\eqnx}{\end{eqnarray}}
\newcommand{\f}[2]{\frac{#1}{#2}}

\newcommand{\sx}{\sum_{i=1}^{N_f}}
\newcommand{\px}{\prod_{i=1}^{N_f}}

\newcommand{\Lmh}{\hat{\Lambda}}
\newcommand{\Qt}{\tilde{Q}}
\newcommand{\Xt}{\tilde{X}}

\newcommand{\Lm}{\Lambda}

\newcommand{\nn}{{\cal N}}
\newcommand{\uf}{u^{fact.}}

\newcommand{\upnc}{{\cal U}^{pure}}
\newcommand{\upm}{{\cal U}^{matter}}
\newcommand{\bin}[2]{\left(\!\!%
\begin{tabular}{c}
{$#1$}\\
{$#2$}
\end{tabular}
\!\!\right)
}
\begin{document}

\begin{titlepage}

\rightline{KUL-TF-2003-18}
\rightline{NORDITA-HE-2003-45}
\setcounter{page}{0}

\vskip 1.6cm

\begin{center}

{\LARGE Affleck-Dine-Seiberg from Seiberg-Witten}

\vskip 1.4cm

Yves Demasure
\\ 
\vskip 0.3cm
 
{\small
Instituut voor Theoretische Fysica, Katholieke Universiteit Leuven,}\\
{\small Celestijnenlaan 200D, B-3001 Leuven, Belgium}\\
{\small  NORDITA,}\\
{\small Blegdamsvej 17, DK-2100 Copenhagen, Denmark}\\

{\small {\tt demasure@nordita.dk}}

\vskip 1.4cm
\end{center}

\begin{abstract}
Perturbing the Seiberg-Witten curves for $\nn=2$ $U(N_c)$ and $SU(N_c)$
super Yang-Mills theory with $N_f<N_c$ flavours with
a mass term for the adjoint field completely lifts the quantum vacuum 
degeneracy. The generated $\nn=1$ effective  superpotential can be 
obtained from the factorization formulae of Seiberg-Witten
curves with matter. We show that the 
Affleck-Dine-Seiberg superpotential emerges. 
Moreover it appears 
additive with respect to the classical superpotential for the 
meson superfields, as expected from the Intriligator-Leigh-Seiberg
linearity principle.
\end{abstract}

\end{titlepage}

\section{Introduction}

Since the Dijkgraaf-Vafa conjecture \cite{DV} there has been an
increased activity in the field of $\nn=1$ gauge theories.
In this note we want to connect recent results 
concerning factorization of Seiberg-Witten curves obtained by using
Dijkgraaf-Vafa with the Affleck-Dine-Seiberg superpotential,
derived from symmetries, anomalies, holomorphicity and an 
instanton calculation \cite{ADS} \cite{is}.

We focus on $\nn=1$ $U(N_c)$ and $SU(N_c)$ 
super Yang-Mills theory (SYM) with a chiral superfield $\Phi$ in 
the adjoint and 
$N_f<N_c$ massive flavours, consisting of chiral superfields
$Q_i$ and $\Qt_i$, $i=1..N_f$, transforming respectively in the fundamental and 
antifundamental of the gauge group. We consider the tree level superpotential: 
\eq \label{tree}
W_{tree} = \frac{M}{2} Tr \Phi^2 + \sum_{i=1}^{N_f} 
\Qt_i  \Phi Q_i + \sum_{i=1}^{N_f} \Qt_i m_i Q_i, 
\eqx
As the superpotential is quadratic in $\Phi$, one can integrate it out
exactly at the  classical level. 
This results in  a $\nn=1$ SYM theory with $N_f$ flavours
and a classical superpotential:
\eq \label{clas}
W_{class} (X_{ij}=\Qt_i Q_j ; M, m_i)
\eqx
whose exact form depends on whether one considers
$U(N_c)$ or $SU(N_c)$.

The non-renormalization theorems state that there are no perturbative 
quantum corrections to this superpotential. 
Though there are non--perturbative
corrections generated by the dynamics of the SYM theory with flavours.
These effects are captured by adding to the classical
superpotential (\ref{clas}) the Affleck-Dine-Seiberg
superpotential (ADS): 
\eq
W_{ADS} = (N_c-N_f) \left(\frac{\Lmh^{3N_c-N_f}}{\det X_{ij}} 
\right)^{\frac{1}{N_c-N_f}}
\eqx
where $\Lmh$ is the dynamically generated scale of the 'downstairs'
SYM theory with $N_f$ flavours. It is related to the scale $\Lm$ of the
'upstairs' theory before integrating out the adjoint by:
\eq \label{mos}
\Lmh^{3 N_c - N_f} = M^{N_c}\Lm^{2 N_c - N_f}
\eqx
The ADS potential is independent of the parameters appearing in the 
classical tree level potential (\ref{clas}). 

The purpose of this note is to obtain this non-trivial structure of
the quantum effective superpotential 
and in particular the Affleck-Dine-Seiberg superpotential
from the factorization
of Seiberg-Witten curves with matter \cite{yr2}.
%The tree level
%superpotential (\ref{tree}) can be interpreted as a $\nn=2$ SYM 
%theory with $N_f$ flavours
%broken to $\nn=1$ by a mass term for the adjoint. 

In \cite{ferretti} agreement between the matrix model calculation
and a field theory calculation using
the ADS potential has been shown for $U(2)$ super Yang-Mills 
with $1$ flavour.
Afterwards \cite{yr1} showed that the ADS potential in terms of the
meson superfield can be obtained using Random Matrix Model techniques {\`a} la 
Dijkgraaf-Vafa. It appeared from the large $N$ limit of the Jacobian 
arising from changing integration variables from $Q_i$ and $\Qt_i$ 
to the gauge invariant integration variable $X_{ij}$. 
The Jacobian was calculated by means of constrained matrix integrals. 
Recently in \cite{jr} there was some progress towards a diagrammatic 
derivation of the Veneziano-Yankielowicz-Taylor superpotential.

The  plan of this note is as follows:
in section 2 we briefly sketch the original derivation of ADS superpotential.
The computation of the effective superpotential by integrating out $\Phi$
and adding ADS can be found in section 3. Factorization formulae of
Seiberg-Witten curves are collected in section 4. From these
formulae we compute in section 5 the effective superpotential and show
how the Affleck-Dine-Seiberg superpotential emerges. 
We close the note with a discussion.

\section{Affleck-Dine-Seiberg potential}

In this section we will briefly sketch the
derivation of the Affleck-Dine-Seiberg potential in the case of 
$\nn=1$ SYM with $N_f<N_c$ massless flavours. 
An excellent review is \cite{is}.

Using the global $U(N_f)_L \times U(N_f)_R \times U(1)_R$ symmetry of the
theory, anomaly considerations and holomorphicity one can show that the
unique form of the non-perturbative superpotential, consistent with
all symmetries is given by:
%and the fields transform as:
%\begin{equation}
%\begin{array}{c | c c c c c}
%                        &    SU(N_f)_L & SU(N_f)_R  & U(1)_B  &  U(1)_A  &     U(1)_R \\
%   (Q_i , \Qt_i)    &(N_f,1) & (1,\bar{N_f})   & (1,-1)     &   (1,1)  &
%                            (\frac{N_f-N_c}{N_f} , \frac{N_f-N_c}{N_f} )\\
%   \lambda_{\alpha}          &    0  & 0    &  0    &    0        &   1      \\
%    \Lmh^{3N_c-N_f}        &  0  & 0 & 0 &  2N_f    &     0       \\
%    det_{ij} Q_i \bar{Q}_j &  1 & 1   &  0      &  2N_f     &  2N_f-2N_c  \\
%    W_{eff}    & 1 & 1 & 0 & 0 & 2 
%\end{array}
%\end{equation}   
%
%$U(1)_R$ is an R-symmetry so the squarks $(\psi_Q, \psi_{\Qt})$ transform with charges
%$(-\frac{N_c}{N_f},-\frac{N_c}{N_f})$. As the gauginos $\lambda_{\alpha}$
%transform with charge $1$, the anomaly produced by the fermion zero modes of the 
%squarks cancel the anomaly from the gauginos. Hence $U(1)_R$ is anomaly free. 
%The only symmetry which is anomalous is the chiral $U(1)_A$ symmetry.
%One can promote it to a genuine symmetry by assigning
%the correct charge to the dynamically generated scale 
%$\Lmh^{3N_c-N_f}=\mu^{3N_c-N_f}exp(2\pi i \tau)$ .
%There is a unique superpotential which is compatible with the required 
\eq
W_{eff} = C(N_c,N_f) \left(\frac{\Lmh^{3N_c-N_f}}{\det X_{ij}} 
\right)^{\frac{1}{N_c-N_f}},
\eqx
where $\Lmh$ is the dynamical scale of the theory.

By giving a large expectation value to one of the flavours 
$<Q_{k}> = <\Qt_{k}>=a_{k}$ one obtains a relation
between $C(N_c,N_f)$ and $C(N_c-1,N_f-1)$. 
%spontaneously breaking $SU(N_c) $ with $N_f$ flavours to  $SU(N_c - 1)$ 
%with $N_f - 1$ flavors at an energy $a_{N_f}$ by the Higgs mechanism. 
By adding a mass term for one of the massless flavours one obtains a 
relation between $C(N_c, N_f)$ and $C(N_c,N_f-1)$.
Consistency requirements amongst these relations
reduces $C(N_c,N_f)$ to a constant $C$:
\eq
C(N_c,N_f) = (N_c-N_f) C^{\frac{1}{N_c-N_f}}
\eqx
The constant $C$ can then be computed via an one instanton calculation
for $N_f=N_c-1$. The non-abelian gauge group is then completely higgsed
and the instanton computation is reliable. 
Detailed analysis for $N_c=2$ and $N_f=1$
reveals that $C=1$ in the $\overline{DR}$ scheme, which concludes the derivation.

It is a very non-trivial statement that the non-perturbative
effects in the presence of a tree level superpotential for the 
meson superfields $W_{tree}(X_{ij})$ are captured by merely
adding the same ADS superpotential to $W_{tree}(X_{ij})$ \cite{ILS}.
The form of the quantum generated superpotential, consistent
with the symmetries is then not anymore unique. There are other possible
terms that one can write down involving the parameters of the classical
potential. Though one can show, using some appropriate limiting
procedures that the only consistent quantum effective superpotential
is given by:
\eq
W_{eff} (X_{ij}, \Lmh) = W_{class} (X_{ij}) +
(N_c-N_f) \left(\frac{\Lmh^{3N_c-N_f}}{\det X_{ij}} 
\right)^{\frac{1}{N_c-N_f}}
\eqx

%The low energy theory is than $SU(N_c)$ with $N_f-1$ flavours.
%Consistency of the family effective superpotentials reduces all $C(N_c,N_f)$
%numbers to one universal co

\section{Superpotential from integrating out $\Phi$}

Starting point is the tree level superpotential (\ref{tree}):
\eq
W_{tree} = \frac{M}{2} Tr \Phi^2 + \sum_{i=1}^{N_f} 
\Qt_i  \Phi Q_i + \sum_{i} \Qt_i m_{i} Q_i, \nonumber
\eqx
As the results for integrating out the superfield
in the adjoint of $U(N_c)$ and $SU(N_c)$ are slightly different,
we discuss them seperately. Note that the ADS non-perturbative
potential has the same form for both cases.

\subsection*{$U(N_c)$ with $N_f$ flavours}

Integrating out $\Phi_{ab}$ in the adjoint of $U(N_c)$ leads straightforward
to:
\eq
W_{class} = - \frac{1}{2 M} Tr X^2 + Tr mX,
\eqx
where $m$ is the diagonal $N_f \times N_f$ mass  matrix diag$(m_1,..,m_{N_f})$
and $X_{ij}=\Qt_i Q_j$. 
The non-perturbative dynamics of $U(N_c)$ SYM theory with $N_f$ flavours
is captured by the addition of the ADS potential:
\eq
W_{eff}(X, M, m_i, \Lmh) = - \frac{1}{2 M} Tr X^2  + Tr mX
 + (N_c - N_f) \left( \frac{\Lmh^{3 N_c - N_f}}{\det X_{ij}} 
\right)^{\frac{1}{N_c -N_f}}
\eqx
Note that $\Lmh$ is the dynamical scale of the 'downstairs' theory
after integrating out $\Phi$. Invoking the matching of the scales 
(\ref{mos}) and  the field redefinition $X=M \Xt$ one can rewrite the
superpotential as:
\eq \label{u1}
W_{eff}(\Xt,M,m_i,\Lm) = M\left(- \frac{1}{2} Tr \Xt^2 +  Tr m\Xt
 + (N_c - N_f) \left( \frac{\Lm^{2 N_c - N_f}}{\det \Xt_{ij}} 
\right)^{\frac{1}{N_c -N_f}}\right)
\eqx
Note that upon using the scale $\Lm$ of the theory with adjoint, 
the superpotential is 
linear in its mass  $M$. This is a consequence 
of the Intriligator-Leigh-Seiberg linearity principle \cite{ILS}.

The final step consists of integrating out the modified meson superfield 
$\Xt_{ij}$ from the superpotential (\ref{u1}). 
All field equations are solved by a diagonal meson matrix
$\Xt=$diag$(\Xt_{11},..,\Xt_{N_fN_f})$. The superpotential and the
equations of motion of the diagonal elements reduce to:
\eqn \label{unp}
W_{eff} &=& M\left(-\frac{1}{2}\sx\Xt_{ii}^2 
+\sx m_i \Xt_{ii}
+(N_c-N_f) \left( \frac{\Lm^{2 N_c - N_f}}{\px\Xt_{ii}} 
\right)^{\frac{1}{N_c -N_f}}\right)  \nonumber  \\
-\Xt_{kk}^2  &+& m_k \Xt_{kk} - 
\left( \frac{\Lm^{2N_c-N_f}}{\px \Xt_{ii}} 
\right)^{\frac{1}{N_c-N_f}} = 0, \quad k=1..N_f
\eqnx
These equations define implicitly the quantum 
effective superpotential
in terms of the flavour masses and the scale $\Lm$. 
It is linear in the mass of the adjoint superfield.

\subsection*{$SU(N_c)$ with $N_f$ flavours}

Integrating out $\Phi_{ab}$ in the adjoint of $SU(N_c)$ 
is a little more subtle: 
\eq
W_{class} = - \frac{1}{2 M} Tr X^2 + \frac{(Tr X)^2}{2 N_c M} + Tr mX
\eqx
where $m$ is again the diagonal $N_f \times N_f$ mass matrix.
Addition of the ADS superpotential encodes all non-perturbative effects:
\eq
W_{eff}(X, M, m_i, \Lmh) = - \frac{1}{2 M} Tr X^2 + 
\frac{(Tr X)^2}{2 N_c M} + Tr mX
 + (N_c - N_f) \left( \frac{\Lmh^{3 N_c - N_f}}{\det X_{ij}} 
\right)^{\frac{1}{N_c -N_f}}
\eqx
This superpotential is linear in $M$, when written in terms of the
scale of the upstairs theory: 
\eq \label{su1}
W_{eff}= 
M \left( - \frac{1}{2} Tr \Xt^2 + \frac{(Tr \Xt)^2}{2 N_c} + Tr m \Xt
 + (N_c - N_f) \left( \frac{\Lm^{2 N_c - N_f}}{\det \Xt_{ij}} 
\right)^{\frac{1}{N_c -N_f}} \right)
\eqx
Integrating out the modified meson superfields $\Xt_{ij}=X_{ij}/M$ 
gives rise to an implicite expression for 
the superpotential $W_{eff}(M, m_i,\Lm)$:
\eqn \label{sunp}
W_{eff}&=&M\left(\sx \frac{-\Xt_{ii}^2}{2} +m_i \Xt_{ii} +
\frac{\big( \sx \Xt_{ii} \big)^2}{2 N_c} 
 + (N_c-N_f) \left( \frac{\Lm^{2N_c-N_f}}{\px \Xt_{ii}} 
\right)^{\frac{1}{N_c-N_f}} \right) \nonumber \\
 -\Xt_{kk} &+& \frac{1}{N_c} \sx \Xt_{ii} + m_i - 
\frac{1}{\Xt_{kk}} \left( \frac{\Lm^{2 N_c - N_f}}{\px \Xt_{ii}} 
\right)^{\frac{1}{N_c -N_f}} = 0, \quad \quad k=1..N_f
\eqnx

\section{Factorization of Seiberg-Witten curves with matter}

In the remaining sections  
we derive the effective
superpotential from the factorization formulae 
of Seiberg-Witten curves with matter. 
We show that the Affleck-Dine-Seiberg 
superpotential emerges with all correct coefficients in addition
to the classical superpotential for the meson superfields.

$\nn=2$ $U(N_c)$ SYM theory with $N_f<N_c$ flavours 
has a quantum Coulomb branch parametrized by $u_p=Tr\frac{\Phi^p}{p}$.
At each point of this $N_c$ dimensional vacuum manifold the low 
energy effective field theory is an abelian $U(1)^{N_c}$ theory.
All the relevant quantum dynamics can be recast at each point 
of the moduli space in 
terms of the Seiberg-Witten curve \cite{SW}\cite{SW2}:
\eq
y^2=P_{N_c}(x,u_k)^2 -4\Lm^{2N_c-N_f} \prod_{i=1}^{N_f} (x+m_i) 
\eqx
It is a family of genus $N_c-1$ hyperelliptic curves parametrized by
$u_p$.

Perturbing  this $\nn=2$ theory with a tree level mass term $M$  for $\Phi$
completely lifts the vacuum degeneracy. An effective potential 
given by \cite{SW}\cite{CIV} \cite{FF}:
\eq \label{pot}
W_{eff}(R,T) = M \uf_2(u_1, \Lm, m_i) 
\eqx
is generated. Note that this effective superpotential is manifest linear in 
$M$. All non-trivial information is encoded in $\uf_2$. It represents
the tuning of the $u_2$ parameter in the Seiberg-Witten curve
such that it completely factorizes:
\eq
y^2=P_{N_c}(x,u_k)^2 -4\Lm^{2N_c-N_f} \prod_{i=1}^{N_f} (x+m_i) 
= F_2(x) H_{N_c-1}^2(x),
\eqx
where the subscript on $P$,$F$ and $H$ denote the degree of the polynomial.
This tuning corresponds to a restriction of the original Coulomb
branch to a $1$ dimensional submanifold where all $N_c-1$ mutually local
monopoles become massless.

These factorization formulae have been found recently using 
Random Matrices techniques in the spirit of Dijkgraaf-Vafa \cite{yr2} 
and give an expression for the moduli 
$u_p$ in terms of one parameter $u_1$, the scale of the
theory $\Lm$ and  the flavour masses $m_i$:
\eq
\uf_p = N_c\, \upnc_p(R,T) +\sum_{i=1}^{N_f}\, \upm_p(R,T,m_i)
\eqx
where
\eqn
\upnc_p(R,T) &=&
 \f{1}{2}  \sum_{q=0}^{[p/2]}
\bin{p}{2q} \bin{2q}{q} R^{q} T^{p-2q-1}\\
\upm_{p\geq 2}(R,T,m)\!\!\! &=&\!\!\! \sum_{n=0}^{p-2} c_{p,n} R
f_n(z) - v_p \f{1}{2} \left( m+T -\sqrt{(m+T)^2-4R} \right)
\eqnx
In the above formula the coefficients $c_{p,n}$ and $v_p$ are given by:
\eqn
c_{p,n} &=& 2^n R^{\f{n}{2}} \sum_{k=0}^{\left[ \f{p-n-2}{2} \right]}
\bin{2k}{k} \bin{p-1}{2k+n+1} R^k T^{p-n-2-2k} \\
v_p &=&  \sum_{q=0}^{[p/2]} \frac{p-2q}{p} 
\bin{p}{2q} \bin{2q}{q} R^{q} T^{p-2q-1}. 
\eqnx
Finally $f_n(z)$ are functions of 
$z=\f{m+T}{2R} \left(m+T+\sqrt{(m+T)^2-4R}\right)$. For the purpose of this
note $f_0(z)=(2z - 2)^{-1}$ will be sufficient. We refer to the original paper 
\cite{yr2} for the explicit expressions for some of the other functions.

%\begin{table}[thb]
%\begin{tabular}{ll}
%$ f_0(z)=\f{1}{2(z-1)} $ & 
%$ f_1(z)=\f{3z-4}{6(z-1)^{3/2}}$ \\
%$ f_2(z)=\f{1}{16(z-1)^2}$ &
%$ f_3(z)=\f{30z^2-65z+32}{120(z-1)^{5/2}}$ \\
%$ f_4(z)=\f{-3z^2+9z-5}{96(z-1)^3}$ &
%$ f_5(z)=\f{525z^3-1610z^2+1582z-512}{3360(z-1)^{7/2}}$ \\
%$ f_6(z)=\f{-48z^3+168z^2-176z+59}{1536(z-1)^4}$ &
%$ f_7(z)=\f{4410z^4-17640z^3+25956z^2-16857z+4096}{40320(z-1)^{9/2}}$
%\end{tabular}
%\caption{Examples of the functions $f_n(z)$ for small $n$.}
%\end{table}

All formulae are in terms of two parameters $R$ and $T$, 
expressing the $\Lm$ and $u_1$ dependence:
\eqn \label{rt1}
u_1(R,T,m) &=& N_cT  -\sx\f{1}{2} \left( m_i+T -\sqrt{(m_i+T)^2-4R} \right)
\\ \label{rt2}
\Lambda^{2N_c - N_f} &=& R^{N_c-N_f} \prod_{i=1}^{N_f}
       \f{1}{2}\left(m_i+T - \sqrt{(m_i + T)^2-4R}\right)
\eqnx

\section{Affleck-Dine-Seiberg from factorized curves}

Using the above formulae one can easily write down the effective
potential (\ref{pot}) where $\uf_2$ is given by:
\eq
 \uf_2=N_c (\frac{1}{2} T^2 - R) +
 \frac{1}{4} \sum_{i=1}^{N_f} \left( (m_i - T)(m_i + T - 
\sqrt{(m_i+T)^2-4R}) -2 R \right)
\eqx 
where $R$ and $T$ are related to $u_1$ and $\Lm$ by (\ref{rt1}) and 
(\ref{rt2}). As all non-trivial information is encoded in $\uf_2$,
we will denote it as the superpotential, keeping in mind that one
still has to multiply everything with $M$.

Defining $x_i = \frac{1}{2} \sqrt{(m_i + T)^2-4R}$ for $i = 1...N_f$, one
can easily eliminate $R$ and $T$ and obtain an implicit expression for the
superpotential: 
\eq
\uf_2 (u_1, x_i, \Lm) = \frac{u_1^2}{2 N_c} + \frac{u_1}{2N_c} \sum_{i=1}^{N_f}
x_i + \frac{1}{2} \sx m_i x_i  + (N_c - \frac{1}{2} N_f) 
\left(\frac{\Lm^{2N_c-N_f}}{\px x_i} \right)^{\frac{1}{N_c-N_f}},
\eqx
where the $x_i$ are functions of $u_1$ and $\Lm$ obtained by
their definition:
\eq
\frac{x_k}{N_c}  u_1 = 
x_k^2  - \frac{1}{N_c} \left( \sx x_i \right) x_k  -  m_k x_k +
\left(\frac{\Lm^{2N_c-N_f}}{\px x_i} \right)^{\frac{1}{N_c-N_f}},
 \quad k=1...N_f
\eqx
Summing these constraints over $k$ gives:
\eq \label{csum}
\frac{u_1}{N_c} \sx x_i  = \sx x_i^2  - 
\frac{1}{N_c} \left( \sx x_i \right)^2  - \sx m_i x_i +
N_f \left(\frac{\Lm^{2N_c-N_f}}{\px x_i} \right)^{\frac{1}{N_c-N_f}}.
\eqx

Note that this form of the superpotential arises merely
from a rewriting of the factorization formulae.

\subsection*{$SU(N_c)$ with $N_f$ flavours}

The results for $SU(N_c)$ super Yang-Mills theories are easily
obtained by setting explicitly $u_1=0$. The superpotential
reduces to:
\eqn
W_{eff} =M \left( \frac{1}{2} \sx m_i x_i  + (N_c - \frac{1}{2} N_f) 
\left(\frac{\Lm^{2N_c-N_f}}{\px x_i} \right)^{\frac{1}{N_c-N_f}} 
\right)  \nonumber\\
x_k^2 - \frac{1}{N_c} \left( \sx x_i \right) x_k - m_k x_k +
\left(\frac{\Lm^{2N_c-N_f}}{\px x_i} \right)^{\frac{1}{N_c-N_f}} = 0,
\quad k=1..N_f
\eqnx
This result is identical as the previous one obtained from the ADS 
superpotential (\ref{sunp}). 
To make the agreement manifest one can combine the 
superpotential with the constraint (\ref{csum}): 
\eqn
W_{eff} &=&M \left(\sx -\frac{x_i^2}{2} + m_i x_i+ 
\frac{\big(\sx x_i \big)^2}{2 N_c} +(N_c-N_f) 
\left(\frac{\Lm^{2N_c-N_f}}{\px x_i} \right)^{\frac{1}{N_c-N_f}} 
\right)  \nonumber\\
x_k^2 &-&\frac{1}{N_c}\left(\sx x_i\right)x_k-m_k x_k +
\left(\frac{\Lm^{2N_c-N_f}}{\px x_i} \right)^{\frac{1}{N_c-N_f}}=0
\quad  k=1..N_f 
\eqnx
This complete form, including the Affleck-Dine-Seiberg superpotential
arises from the factorization formulae of Seiberg-Witten curves.
Notice that we not only get the same endresult, but that the
complete structure of integrating out the meson superfields
is present upon choosing the 'good' variables to rewrite
the factorization formulae. The classical superpotential for the
meson superfield, as well as the ADS superpotential is manifest 
with all the correct coefficients.

\subsection*{$U(N)$ with $N_f$ flavours}

In this case the superpotential is an implicit function of $u_1$
and one need to integrate it out. Minimizing the superpotential
with respect to $u_1$ while using the constraints gives:
\eq \label{min}
\frac{d}{d u_1}W_{eff}(M, m_i, \Lm, u_1, x_k) = 
\frac{u_1}{N_c} + \sx \frac{\partial W_{eff}(M, m_i, \Lm, u_1, x_k)}
{\partial x_i}\frac{\partial x_i}{\partial u_1}=0
\eqx
The last part can be calculated by using:
\eq
\frac{\partial x_i}{\partial u_1} = \left( 
\frac{\partial u_1}{\partial x_i}\right)^{-1} \left( 1 - \sum_{j \neq i}
\frac{\partial u_1}{\partial x_j}\frac{\partial x_j}{\partial u_1} \right)
\eqx
By manipulating carefully the equations  we obtain:
\eq
 \frac{\partial W_{eff}}{\partial x_i}
\left( \frac{\partial u_1}{\partial x_i}\right)^{-1} = \frac{1}{N_c} 
\sum_{k=1}^{N_f} x_k
\eqx
As this expression is independent of the label $i$ the total result
(\ref{min}) simplifies to:
\eq
\frac{u_1}{N_c} + \frac{\sum_{k=1}^{N_f} x_k}{N_c} \sx  \left( 
1 - \sum_{j \neq i}
\frac{\partial u_1}{\partial x_j}\frac{\partial x_j}{\partial u_1} \right)
= 0
\eqx
The remaining summation is trivial and
the value for $u_1$ minimizing the potential is given by
\eq
u_1 = - \sx x_i
\eqx
The final step consists of plugging this value for $u_1$ into the 
superpotential:
\eqn
W_{eff}(M, m_i, \Lm, x_i) &=& M \left(\frac{1}{2} m_i x_i
+ (N_c - \frac{1}{2}N_f)  
\left(\frac{\Lm^{2N_c-N_f}}{\px x_i} \right)^{\frac{1}{N_c-N_f}} 
\right) \nonumber \\
x_k^2 &-& m_k x_k + 
\left(\frac{\Lm^{2N_c-N_f}}{\px x_i} \right)^{\frac{1}{N_c-N_f}} = 0
\quad \quad k=1..N_f
\eqnx
Proceeding as above and using the constraint (\ref{csum})
one can make the agreement  with the result 
using the explicit form of ADS(\ref{unp}) manifest:
\eqn
W_{eff} &=& M \left(-\frac{1}{2} \sx x_i^2 + \sx m_i x_i
+ (N_c-N_f)  
\left(\frac{\Lm^{2N_c-N_f}}{\px x_i} \right)^{\frac{1}{N_c-N_f}} 
\right) \nonumber \\
x_k^2 &-& m_k x_k + 
\left(\frac{\Lm^{2N_c-N_f}}{\px x_i} \right)^{\frac{1}{N_c-N_f}} = 0
\quad \quad k=1..N_f
\eqnx
Again the ADS superpotential arises additive with respect to
the classical superpotential for the meson superfields as expected.
Moreover the form of the effective superpotential mimics the integrating 
out of the meson superfield from the method involving the ADS superpotential.

\subsection*{Discussion}

Perturbing the $\nn=2$ Seiberg-Witten curve with $N_f$ flavours with
a mass term for the adjoint breaks the supersymmetry to $\nn=1$. 
We computed the effective superpotential from the
factorization formulae of the Seiberg-Witten curves and 
showed that the Affleck-Dine-Seiberg superpotential emerges. It appears
additive with respect to the classical superpotential for the
meson superfields as expected from the Intriligator-Leigh-Seiberg
linearity principle. Moreover the complete structure of integrating
out the meson fields can be retrieved from the factorization formulae.

\bigskip

\noindent{\bf Acknowledgments} 
It is a pleasure to thank Paolo Di Vecchia, Frank Ferrari and 
Paolo Merlatti for discussions.
This work is supported in part by the Federal Office for Scientific, 
Technical and Cultural Affairs through the Interuniversity Attraction 
Pole P5/27 and in part by the European Community's Human Potential 
Programme under contract HPRN-CT-2000-00131 Quantum Spacetime and 
by an EC Marie Curie Training site Fellowship at Nordita, under 
contract number HPMT-CT-2000-00010.

\end{document}